\documentstyle[epsfig]{mn}

\newif\ifAMStwofonts

\def\gtsima{$\; \buildrel > \over \sim \;$}
\def\ltsima{$\; \buildrel < \over \sim \;$}
\def\gsim{\lower.5ex\hbox{\gtsima}}
\def\lsim{\lower.5ex\hbox{\ltsima}}
\newcommand{\etal}{et al.\ }
\def\Msun{{M_\odot}}
\def\Zsun{{Z_\odot}}
\def\be{\begin{equation}}
\def\ee{\end{equation}}

\ifoldfss
  \ifCUPmtlplainloaded \else
    \NewTextAlphabet{textbfit} {cmbxti10} {}
    \NewTextAlphabet{textbfss} {cmssbx10} {}
    \NewMathAlphabet{mathbfit} {cmbxti10} {} 
    \NewMathAlphabet{mathbfss} {cmssbx10} {} 
  \fi
  \ifAMStwofonts
    \ifCUPmtlplainloaded \else
      \NewSymbolFont{upmath} {eurm10}
      \NewSymbolFont{AMSa} {msam10}
      \NewMathSymbol{\upi}     {0}{upmath}{19}
      \NewMathSymbol{\umu}     {0}{upmath}{16}
      \NewMathSymbol{\upartial}{0}{upmath}{40}
      \NewMathSymbol{\leqslant}{3}{AMSa}{36}
      \NewMathSymbol{\geqslant}{3}{AMSa}{3E}

      \let\leq=\leqslant 
      \let\geq=\geqslant 
    \fi
  \fi
\fi 

\ifnfssone
  \newmathalphabet{\mathit}
  \addtoversion{normal}{\mathit}{cmr}{m}{it}
  \addtoversion{bold}{\mathit}{cmr}{bx}{it}
  \newmathalphabet{\mathbfit} 
  \addtoversion{normal}{\mathbfit}{cmr}{bx}{it}
  \addtoversion{bold}{\mathbfit}{cmr}{bx}{it}
  \newmathalphabet{\mathbfss} 
  \addtoversion{normal}{\mathbfss}{cmss}{bx}{n}
  \addtoversion{bold}{\mathbfss}{cmss}{bx}{n}
  \ifAMStwofonts
    \ifCUPmtlplainloaded \else
      \UseAMStwoboldmath
      \makeatletter
      \new@mathgroup\upmath@group
      \define@mathgroup\mv@normal\upmath@group{eur}{m}{n}
      \define@mathgroup\mv@bold\upmath@group{eur}{b}{n}
      \edef\UPM{\hexnumber\upmath@group}
      \new@mathgroup\amsa@group
      \define@mathgroup\mv@normal\amsa@group{msa}{m}{n}
      \define@mathgroup\mv@bold\amsa@group{msa}{m}{n}
      \edef\AMSa{\hexnumber\amsa@group}
      \makeatother
      \mathchardef\upi="0\UPM19
      \mathchardef\umu="0\UPM16
      \mathchardef\upartial="0\UPM40
      \mathchardef\leqslant="3\AMSa36
      \mathchardef\geqslant="3\AMSa3E

      \let\leq=\leqslant 
      \let\geq=\geqslant 
    \fi
  \fi
\fi 

\ifnfsstwo
  \DeclareMathAlphabet{\mathbfit}{OT1}{cmr}{bx}{it}
  \SetMathAlphabet\mathbfit{bold}{OT1}{cmr}{bx}{it}
  \DeclareMathAlphabet{\mathbfss}{OT1}{cmss}{bx}{n}
  \SetMathAlphabet\mathbfss{bold}{OT1}{cmss}{bx}{n}
  \ifAMStwofonts
    \ifCUPmtlplainloaded \else
      \DeclareSymbolFont{UPM}{U}{eur}{m}{n}
      \SetSymbolFont{UPM}{bold}{U}{eur}{b}{n}
      \DeclareSymbolFont{AMSa}{U}{msa}{m}{n}
      \DeclareMathSymbol{\upi}{0}{UPM}{"19}
      \DeclareMathSymbol{\umu}{0}{UPM}{"16}
      \DeclareMathSymbol{\upartial}{0}{UPM}{"40}
      \DeclareMathSymbol{\leqslant}{3}{AMSa}{"36}
      \DeclareMathSymbol{\geqslant}{3}{AMSa}{"3E}

      \let\leq=\leqslant 
      \let\geq=\geqslant 
    \fi
  \fi
\fi 

\ifCUPmtlplainloaded \else
  \ifAMStwofonts \else 
    \def\upi{\pi}
    \def\umu{\mu}
    \def\upartial{\partial}
  \fi
\fi

\title{Life and times of dwarf spheroidal galaxies}
\author[Stefania Salvadori, Andrea Ferrara \& Raffaella Schneider]
{Stefania Salvadori$^{1}$, Andrea Ferrara$^{1}$ \& Raffaella Schneider$^{2}$\\
$^1$SISSA/International School for Advanced Studies, Via Beirut 4, 34100 Trieste, Italy\\ 
$^2$INAF/Osservatorio Astrofisico di Arcetri, Largo Enrico Fermi 5, 50125 Firenze, Italy}
\date{}

\pagerange{\pageref{firstpage}--\pageref{lastpage}}
\pubyear{}

\begin{document}

\maketitle 
\label{firstpage}

\begin{abstract}

We propose a cosmological scenario for the formation and evolution
of dwarf spheroidal galaxies (dSphs), satellites of the Milky Way (MW).
An improved version of the semi-analytical code GAMETE (GAlaxy
Merger Tree \& Evolution) is used to follow the dSphs evolution
simultaneously with the MW formation, matching the observed properties
of both. 
In this scenario dSph galaxies represent fossil objects virializing at
$z=7.2\pm 0.7$ (i.e. in the pre-reionization era $z>z_{rei}=6$) in the
MW environment which at that epoch has already been pre-enriched up to 
[Fe/H]$\sim -3$; their dynamical masses are in the narrow range
$M=(1.6\pm 0.7)\times10^8 \Msun$, although a larger spread might be
introduced by a more refined treatment of reionization. Mechanical
feedback effects are dramatic in such low-mass objects, causing the
complete blow-away of the gas $\sim 100$~Myr after the formation
epoch: $99\%$ of the present-day stellar mass, $M_*=(3\pm 0.7)\times
10^6 \Msun$, forms during this evolutionary phase, i.e. their age is
$>13$~Gyr. Later on, star formation is re-ignited by returned gas from
evolved stars and a second blow-away occurs. The cycle continues for
about 1 Gyr during which star formation is intermittent. At $z=0$ the
dSph gas content is $M_g=(2.68\pm 0.97)\times 10^4 \Msun$. 
Our results match several observed properties of Sculptor, used as a
template of dSphs: (i) the Metallicity Distribution Function; (ii) the
Color Magnitude Diagram; (iii) the decrement of the stellar [O/Fe]
abundance ratio for [Fe/H]$>-1.5$; (iv) the dark matter content and
the light-to-mass ratio; (v) the HI gas mass content.

\end{abstract}


\section{Background}

The lack of a comprehensive scenario for the formation and evolution
of dwarf spheroidal galaxies (dSphs) contrast with the large amount of
available data for these nearby Local Group satellites.
Several authors have focused on different aspects of the dSph evolution
and on the observed properties related with them, giving important
contributions to our actual understanding of such puzzling
objects. However, many questions remain unanswered and models able to
match simultaneously different observed properties are still missing. 

Recent observations by Helmi et al. (2006) opened new questions
on the origin of dSph galaxies. By observing the stellar
Metallicity Distribution Function (MDF) in four nearby dSphs, they
found a significant lack of stars with [Fe/H]$<-3$. On the contrary,
the Galactic halo MDF shows a low [Fe/H]-tail extending down to
[Fe/H]$\sim -4$ (Beers \& Christlieb 2006) and even below (Christlieb
et al. 2002, 2006; Frebel et al. 2005, Christlieb 2007). 
Does such result imply that dSph and MW progenitors are different? 
Is the dSph birth environment pre-enriched or does the Initial Mass 
Function (IMF) behave differently in Galactic building blocks and in 
dSphs at earliest times? 
If the pre-enrichment can solve the problem it must be completed
before the beginning of the star formation (SF) in dSphs
.i.e. $>13$~Gyr ago. The dSph star formation histories (SFH) derived
by the analysis of the observed color-magnitude diagram (CMD), in
fact, show that all dSph have an ancient stellar population formed
more that 13~Gyr ago (Grebel \& Gallagher 2004). Which mechanism can
be responsible of such rapid metal-enrichment? 

In addition to these new open questions there is still a crucial
unresolved problem related to the dSph's SFHs. In fact, although all
dSphs display an ancient stellar population, their SFHs appear to be
considerably different. In most of them the bulk of stars is made by
ancient stars ($>10$~Gyr old) i.e. their SF activity is concentrated
during the first Gyrs (Dolphin et al.~2005). However
Fornax, LeoII and Sagittarius show very different features: in these
objects the bulk of the stars was formed much less than 10~Gyr ago and
their SF activity proceeds until $z\sim 1$, or even to lower
redshifts (Grebel \& Gallagher 2004). 
Carina, finally, exhibits a clearly episodic SFH, with a
pause of several Gyrs after the old population formed and a massive
formation of stars younger than 10~Gyr (Smecker-Hane et al.~1994,
Hurley-Keller, Mateo \& Grebel 1999). As pointed out by Grebel \&
Gallagher (2004) such variety of SFHs cannot be explained with a drop in the SF
activity in response to photoionization; only Carina in fact displays
the gap expected if the SF were truncated during reionization and then
restarted. If reionization cannot explain the inferred variety of
SFHs, local processes, such as mechanical feedback, tidal stripping,
gas infall etc., must be invoked in order to produce these observational
features. However, it is difficult that local processes can lead to
different SFHs if the dSph host halo mass has a ''Universal'' value, 
as predicted by Mateo et al. (1998), Gilmore et al. (2007) and Walker et~al. 
(2007), and if the assumed initial gas-to-mass ratio is taken to be equal 
to the cosmic mean value. Will more sophisticated observations and analysis 
reveal a spread in the dSph DM content? If this is the case, different 
host halo mass could be more easily linked to different SFHs by internal 
physical processes. In addition to these issues there are a number of 
related aspects that deserve attention, as for example the observed elemental 
abundance patterns (like the [$\alpha$/Fe] ratio and the s-elements abundance).

Despite all of these unresolved questions, the relevance of stellar
feedback for the evolution of dSph galaxies (Ferrara \& Tolstoy 2002
and references therein) can be considered a milestone of our actual
understanding of these objects, given the general consensus among theoretical
studies and the continuous confirmation by observational evidences.
Given their low DM content and therefore their shallow potential
wells, mechanical feedback driven by SN explosions has dramatic 
effects in dSphs. Such intense mechanical feedback activity is
indirectly confirmed by the low light-to-dark matter ratios observed
in dSphs (Mateo et. al 1998, Gilmore et al. 2007). Moreover,
observations of neutral hydrogen (HI), reveal that the Local Group
dSphs are all relatively HI poor, suggesting that maybe most of the
gas have been removed from these galaxies.  
The Sculptor dSph is one of the few with HI emission; a lower limit of 
$M_{HI}> 3\times 10^4\Msun$ has been derived by Carignan et al. (1998)
using radio observations. In addition, winds are presumably
metal-enhanced, as suggested by several theoretical studies (Vader
1998; MacLow \& Ferrara 1999; Fujita et al. 2004) and by recent X-rays
observations of the starburst galaxy NGC 1569 (Martin, Kobulnicky \&
Heckman 2002).

Many theoretical works have attacked some aspects of the dSph
evolution separately: 
\begin{itemize}
\item{Origin \& DM content of dSphs}: 
  Bullock et~al. (2001); Kravtsov et~al. (2004); Ricotti \& Gnedin (2005); 
  Gnedin \& Kravtsov (2006); Read, Pontzen \& Viel (2006); Moore et al. (2006);
  Metz \& Kroupa (2007).
\item{DSph MDFs}: Ripamonti et al. (2006); Lanfranchi \& Matteucci
  (2007)
\item{SFH \& abundance ratios of dSphs}:  Ikuta \& Arimoto (2002);
  Fenner et al. (2006);  Lanfranchi \& Matteucci (2007);
  Stinson et~al. (2007).
\item{Stellar feedback \& dSph gas content}: Ferrara \& Tolstoy
  (2000); Tassis et.~al (2003); Fujita et al. (2004); Lanfranchi \&
  Matteucci (2007). 
\end{itemize}

In this study we analyze the formation and evolution of dSph galaxies,
satellites of the MW, in their cosmological context by using the
improved GAlaxy MErger Tree \& Evolution code (GAMETE), which allows
to build-up the SFH and chemical enrichment of the MW along its
hierarchical merger tree (Salvadori, Schneider \& Ferrara 2007,
hereafter SSF07). This approach gives a self-consistent description of
the dSphs evolution and MW formation: dSphs form out from their
natural birth environment, the Galactic Medium (GM), whose metallicity
evolution is completely determined by the history of SF and by mechanical
feedback processes along the build-up of the Galaxy. 
The star formation and mechanical feedback efficiencies of dSphs are
assumed to be the same as for all the Galactic building blocks and
calibrated to reproduce the observable properties of the MW. The model
allows to predict the following observable properties of a typical
dSph galaxy: (i) the formation epoch; (ii) the MDF; 
(iii) the SFH and the derived CMD diagram; (iv) the stellar [O/Fe]
abundance ratio with respect to [Fe/H]; (v) the DM content; (vi) the
stellar-to-mass ratio; (vii) the final gas content. These are compared
with observations of Sculptor, which represents the best studied
nearby dSph. A global scenario for the formation and evolution of dSph
galaxies is presented.

The plan of the paper is the following: a recap of the general
properties of the improved code GAMETE is presented in Sec.~2. 
The life of a dSph galaxy is described in subsequent sections: the birth
environment and the selection criteria from the MW building blocks are
presented in Sec.~3; the evolution until the blow-away, driven by
mechanical feedback, is described in Sec.~4; the subsequent and final 
stages of the dSph life are traced in Sec.~5. Sec.~6 is devoted to
a comparison between model results and some observational properties
of the Sculptor dSph. Finally, a summary and discussion of 
the main results is given in Sec.~7.   

\section{Model Description}
In this Section, we first summarize the main features of the model
introduced in SSF07; following that we discuss the modifications and
improvements made for the purpose of this work. 

\subsection{Summary of the model}
The semi-analytic code GAMETE used in SSF07 allows to follow the star 
formation history and chemical enrichment of the MW throughout
its hierarchical merger tree. Its main features can be summarized along
the following points (for detailed explanations see SSF07). The code 
reconstructs the hierarchical merger history of the MW using a
Monte Carlo algorithm based on the extended Press \& Schechter theory 
(Bond et al. 1991; Lacey \& Cole, 1993); it adopts a {\it binary} 
scheme with {\it accretion} (Cole et al. 2000, Volonteri, Haardt 
\& Madau 2003) to decompose the present day MW dark matter halo 
($M_{MW}\sim 10^{12}\Msun$ Binney \& Merrifield 1998) into its 
progenitors, running backward in time up to $z=20$. At any time 
a halo of mass $M_0$ can either loose part of its mass 
(corresponding to a cumulative fragmentation into haloes 
with $M<M_{res}$) or loose mass and fragment into two
progenitor haloes with random masses in the range $M_{res}<M<M_0/2$. 
The mass below the resolution limit accounts for the {\it Galactic Medium} 
(GM) which represents the mass reservoir into which
haloes are embedded. During the star formation history of the MW, the
progenitor haloes accrete material from the GM and virialize out of it.  
We assume that feedback effects rapidly suppress star formation
in the first mini-haloes and that only Ly$\alpha$ cooling haloes
($T_{vir}>10^4$K) contribute to the star formation history and 
chemical enrichment of the Galaxy. This motivates our choice of 
a resolution mass $M_{res}=M_4(z)/10$, where   
\be 
M_4(z)=M(T_{vir}=10^4 {\rm K},z)\sim10^{8}\Msun\left(\frac{10}{1+z}\right)^{3/2}
\label{eq:M4} 
\ee
\noindent
represents the halo mass corresponding to a virial equilibrium
temperature $T_{vir}=10^4$K at a given redshift $z$. At the highest
redshift of the simulation, $z\approx 20$, the gas present in virialized
haloes is assumed to be of primordial composition. The star formation rate
is taken to be proportional to the mass of gas. Following the critical
metallicity scenario (Bromm \etal 2001; Omukai 2000, Omukai \etal 2005; 
Schneider \etal 2002, 2003, 2006; Bromm \& Loeb 2004) we assume that 
low-mass star formation is triggered by the presence of metals 
(and dust) exceeding $Z_{cr}=10^{-5\pm 1}\Zsun$. When the gas in 
star forming haloes has a metallicity $Z\leq Z_{cr}$ Pop~III stars
form with a reference mass value of $m_{popIII}=200\Msun$. If on the
contrary $Z>Z_{cr}$, Pop~II/I stars form according to a Larson 
initial mass function (IMF):
\be
\Phi(m)=\frac{dN}{dm}\propto m^{-1+x}\exp(-m_{cut}/m),
\label{eq:LarsonIMF}
\ee
with $x=-1.35$, $m_{cut}=0.35 \Msun$ and $m$ in the range $[0.1-100]
\Msun$ (Larson 1998). Due to the lack of spatial information,
when metals and gas are returned to the interstellar medium (ISM)
through stellar winds and SN explosions, they are assumed to be 
instantaneously and homogeneously mixed with the ISM. 
The same instantaneous perfect mixing approximation is applied to 
material ejected out of the host halo into the external GM.  

\subsection{New features}

We now discuss the additional features we have incorporated in the model. 
The aim of introducing these new physics is to obtain a more complete 
description of the evolution of a single dSph galaxy. These can be summarized
as follows:
\begin{itemize}
\item {\it Infall rate.} The gas in newly virialized haloes is accreted with an infall rate
given by
\be
\frac{dM_{inf}}{dt}=A\left(\frac{t}{t_{inf}}\right)^2\exp{\left(-\frac{t}{t_{inf}}\right)}.
\label{eq:Minf}
\ee
The selection of this particular functional form has been guided by
the results of simulations presented in Kere\v{s} et al. (2005).
For reasons that will be clarified in Sec.~6.1, the infall time is
assumed to be proportional to the free-fall time, $t_{inf}=t_{ff}/4$ 
where $t_{ff} = (3 \pi/32 G \rho)^{1/2}$, $G$ is the gravitational
constant, and $\rho$ is the total (dark + baryonic) mass density of 
the halo. The normalization constant is set to be 
$A = {2}({\Omega_b}/{\Omega_M}) M/t_{inf}$ so that for $t \rightarrow \infty$
the accreted gas mass reaches the universal value  
$M_{inf}(\infty)=(\Omega_b/\Omega_M) M$. 
No infall is assumed after a merging event i.e. all the
gas is supposed to be instantaneously accreted. Hydrodynamical
simulations in fact show that galaxy mergers can drive significant
inflow of gas raising the star formation rate by more than an order of
magnitude (Mihos \& Hernquist, 1996 and references therein). 

\item {\it Finite stellar lifetimes}.  We follow the chemical evolution
of the gas taking into account that stars of different masses 
evolve on characteristic time-scales (Lanfranchi \& Matteucci 2007). 
The rate at which gas is returned to the ISM 
through winds and SN explosions is computed as:       
\be
\frac{dR(t)}{dt}=\int^{100\Msun}_{m_1(t)}(m-w_m(m))\Phi(m){\mbox SFR}(t-\tau_m)dm,
\ee
where $\tau_m=10/m^2$~Gyr is the lifetime of a star with mass $m$,
$w_m$ is the remnant mass and $m_1(t)$ the turnoff mass i.e. the mass
corresponding to $t=\tau_m$. Similarly, the total ejection rate of an
element ${\it i}$, newly synthesized inside stars (first term in the
parenthesis) and re-ejected into the ISM without being re-processed
(second term), is
\be
\frac{dY_i(t)}{dt}=\int^{100\Msun}_{m_1(t)}\big[(m-w_m(m)-m_i(m,Z))\times
\label{eq:Return}
\ee
\[
Z_i(t-\tau_m)+m_i(m,Z)\big]\Phi(m){\mbox SFR}(t-\tau_m)dm,
\]
\noindent
where $m_i(m,Z)$ is the mass of element ${\it i}$ produced by a star with
initial mass $m$ and metallicity $Z$ and $Z_i(t-\tau_m)$ is the
abundance of the ${\it i-th}$ element at the time $t-\tau_m$. The SN rate is
simply computed as
\be
\frac{dN_{SN}(t)}{dt}=\int^{40\Msun}_{m_1(t) > 8\Msun}\Phi(m){\mbox SFR}(t-\tau_m)dm.
\ee
We used the grid of values $w_m(m)$ and $m_i(m,Z)$ by Heger \& Woosley
(2002) for $140\Msun<m<260\Msun$, Woosley \& Weaver (1995) for
$8\Msun<m<40\Msun$ and van der Hoek \& Groewengen (1997) for
$0.9\Msun<m<8\Msun$.  
\item {\it Mechanical feedback.} Assuming a continuous mass loss prescription (Larson 1974),
the mass of gas ejected into the surrounding GM is regulated by the equation,
\be
\frac{1}{2}M_{ej}v_e^2=E_{SN}
\label{eq:Mej1}
\ee
\noindent
where 
\be
E_{SN}=\epsilon_w N_{SN}\langle E_{SN}\rangle 
\label{eq:Esn}
\ee 
is the kinetic energy injected by SN-driven winds and $v_e^2=GM/r_{vir}=2E_b/M$ is the escape 
velocity of the gas from a halo with mass $M$ and binding energy $E_b$ given by (Barkana \& Loeb 2001)
\[
E_{b}=\frac{1}{2}\frac{GM^{2}}{r_{vir}}= 5.45\times{10^{53}}{\rm erg}
\left(\frac{{M}_{8}}{h^{-1}}\right)^{5/3}\left(\frac{1+z}{10}\right){h}^{-1}.
\]
\label{eq:Eb}
\noindent
In eq.~\ref{eq:Esn}, $\epsilon_{w}$ is a free parameter which controls
the conversion efficiency of SN explosion energy in kinetic form,
$N_{SN}$ is the number of SN, and $\langle E_{SN}\rangle$ is the average
explosion energy; the latter quantity is taken to be equal to 
$2.7\times 10^{52}$erg for Pop~III stars and to 
$1.2\times 10^{51}$erg for Type~II SNe. Differentiating eq.~\ref{eq:Mej1} 
we find that the gas ejection rate
is proportional to the SN explosion rate,  
\be
\frac{dM_{ej}}{dt}=\frac{2 \epsilon_w \langle
E_{SN}\rangle}{v_{e}^2} \frac{dN_{SN}}{dt}.
\label{eq:Mej}
\ee

\item {\it Differential winds.} The gas ejected out of the host halo is 
assumed to be metal-enhanced with respect to the
star forming gas. According to Vader (1986), who
studied SN-driven gas loss during the early evolution of elliptical
galaxies, the SN ejecta suffer very limited mixing before they leave
the galaxy, playing a minor role in its chemical evolution. Such hypothesis
implies different ejection efficiency for gas and metals. This
result has been later confirmed by numerical studies (Mac Low \&
Ferrara, 1999; Fujita et~al. 2004). Adopting a simple prescription,
we assume that the abundance of the ${\it i-th}$ element in the wind 
is proportional to its abundance in the ISM, $Z^w_i=\alpha
Z^{ISM}_i$, and we take $\alpha=10$ only for newly virialized haloes
($M<10^9\Msun$) otherwise $\alpha=1$.  
\end{itemize}

For any star forming halo of the MW hierarchy, we therefore solve
the following system of differential equations:  

\be
\frac{dM_*}{dt} = {\mbox SFR} = \epsilon_*\frac{M_g}{t_{ff}},
\label{eq:SFR}
\ee
\be
\frac{dM_g}{dt} = - {\mbox SFR} + \frac{dR}{dt} + \frac{dM_{inf}}{dt} - \frac{dM_{ej}}{dt},
\label{eq:Mg}
\ee
\be
\frac{dM_{Z_i}}{dt} = - Z^{ISM}_i{\mbox SFR} + \frac{dY_i}{dt} +
Z_i^{vir}\frac{dM_{inf}}{dt} - Z^w_i\frac{dM_{ej}}{dt}.
\label{eq:Mz}
\ee
\noindent
The first equation is the star formation rate; $M_g$ is the
mass of cold gas inside haloes, $\epsilon_*$ the free parameter which
controls the star formation efficiency and $t_{ff}$ the free-fall
time. The second equation describes the mass variation of cold gas: 
it increases because of gas infall and/or returned from stars and 
decreases because of star formation and gas ejection into the GM. 
The third equation, analogous to the second one, regulates the 
mass variation of an element $i$; $Z^{ISM}_i$, $Z_i^{vir}$, and
$Z^{w}_i$ are the abundance of the ${\it i-th}$ element in the ISM,  
in the infalling gas (i.e. in the hot gas at virialization), and
in the wind, respectively.

\begin{figure*}
  \centerline{\psfig{figure=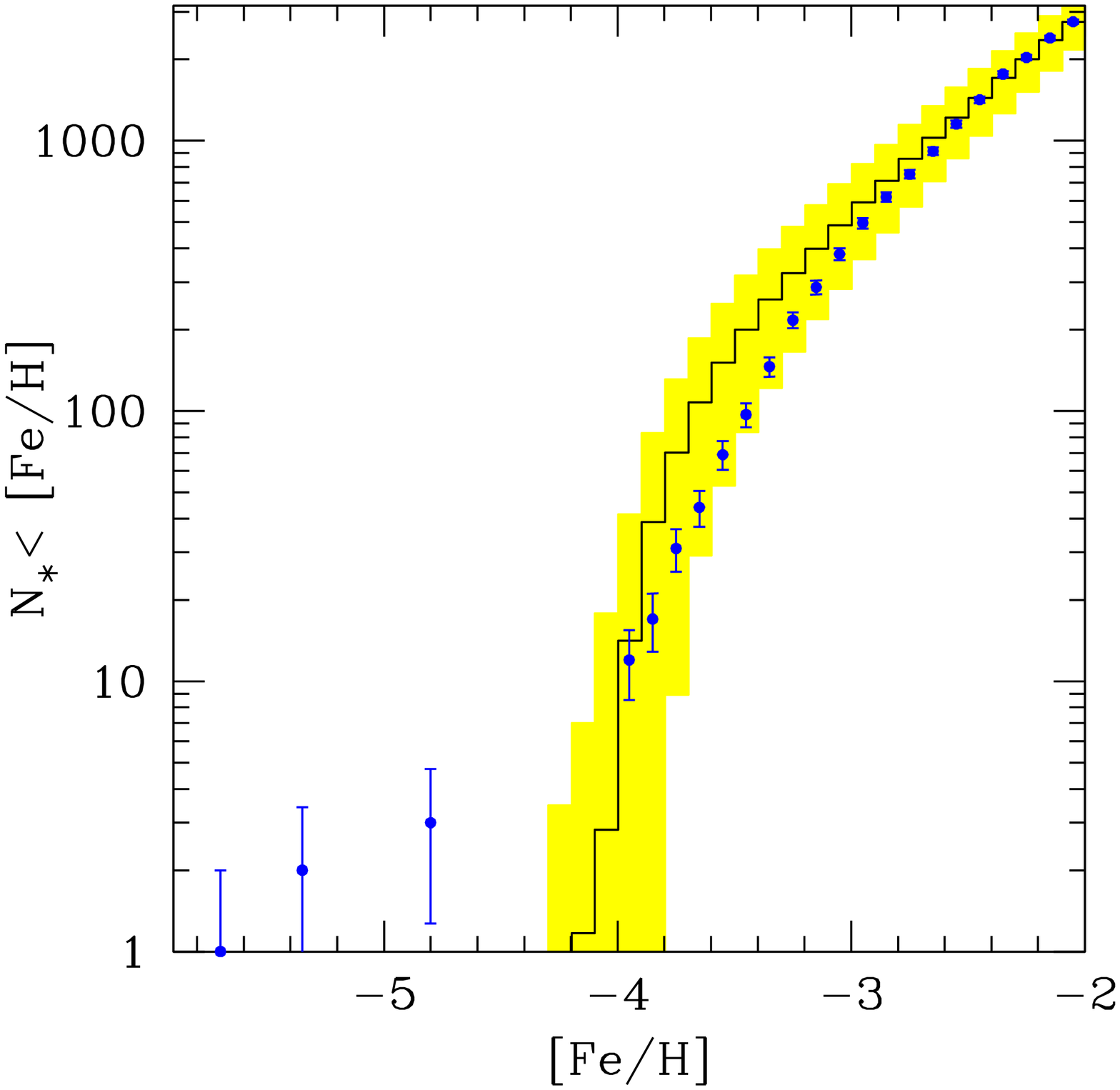,width=5.9cm,angle=0}
    \psfig{figure=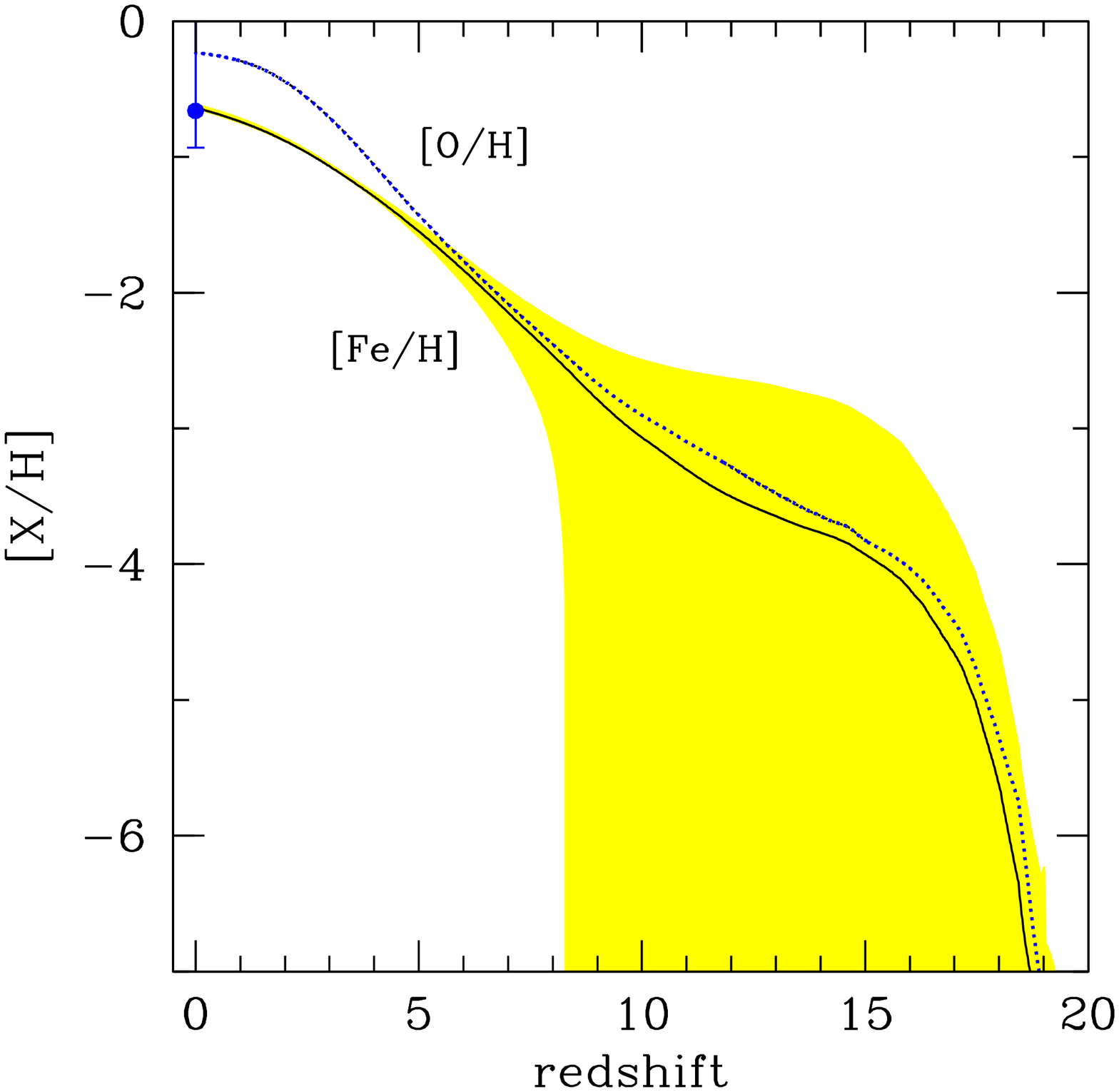,width=5.9cm,angle=0}
    \psfig{figure=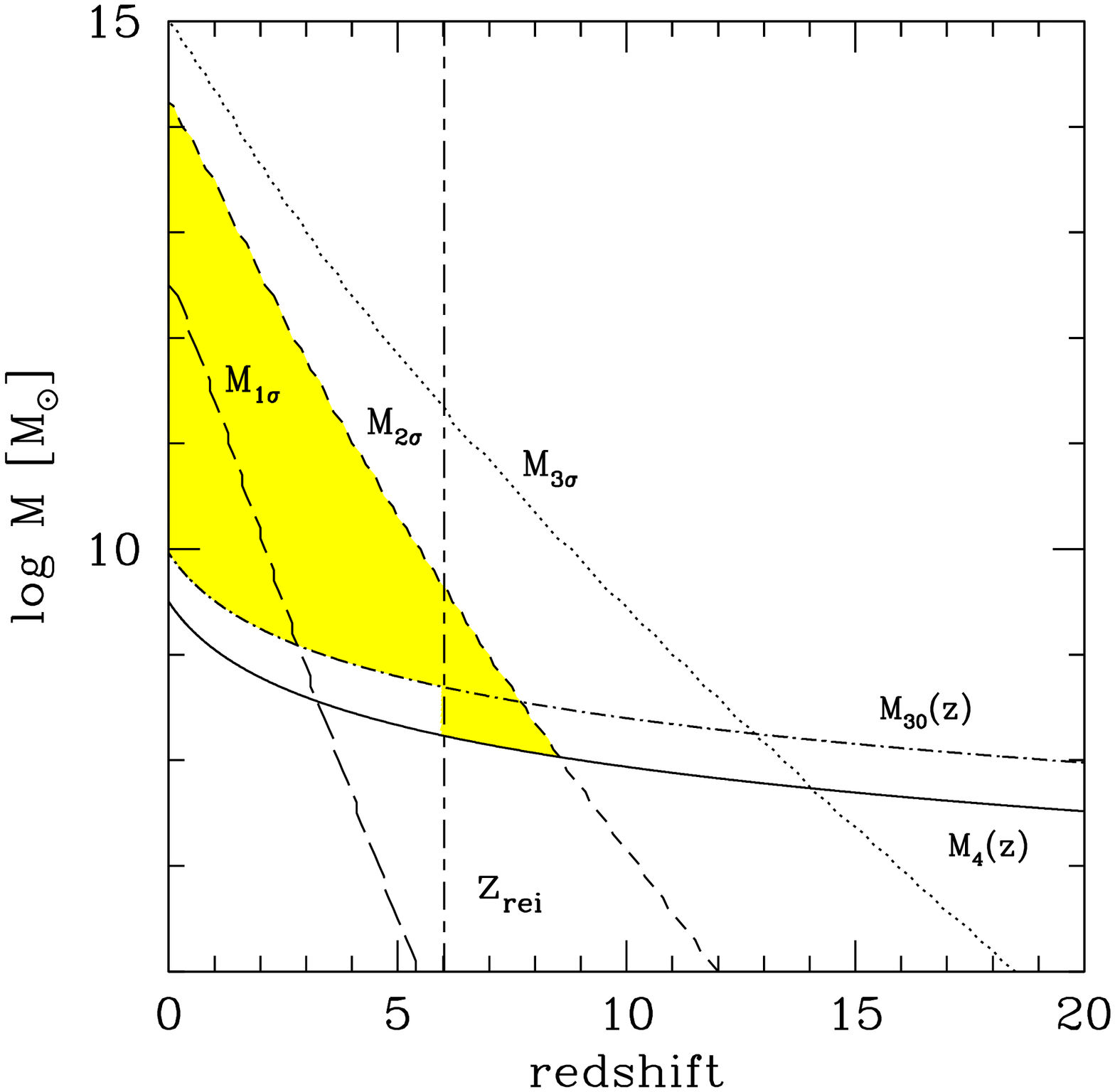,width=5.9cm,angle=0}}
  \caption{{\it Left panel}: Cumulative number of stars below a given
  [Fe/H] observed by Beers \& Christlieb (2006) in the Galactic halo,
  with the inclusion of the three hyper-iron poor stars (Christlieb et
  al. 2002, 2006; Frebel et al. 2005, Christlieb 2007)
  (points), and simulated by using the fiducial model (histogram). 
  The histogram is the average value over
  100 realizations of the merger tree re-normalized to the number of
  observed stars with [Fe/H]$\leq -2$. The shaded area represents $\pm
  1\sigma$ errors. {\it Middle}: Evolution of GM iron (solid line) and
  oxygen (dotted line) abundance. Lines are the average values over
  100 realizations of the merger tree. The shaded area delimits the
  $\pm 1\sigma$ dispersion region for [Fe/H]. The point is the
  measured [O/H] in high-velocity clouds (Ganguly et al. 2005).  
  {\it Right}: Evolution of the mass corresponding to 1 (long dashed
  line), 2 (short dashed line) and 3$\sigma(M,z)$ (dotted line)
  density peaks; the solid line show the evolution of $M_4(z)$ (eq.\ref{eq:M4}), the
  dotted-dashed line the evolution of $M_{30}(z)$. The selected
  reionization redshift $z_{rei}=6$ is also shown in the Figure. 
  The shaded area delimits the region $M_4(z)<M<M_{2\sigma}$ for
  $z>z_{rei}$, $M_4(z)<M<M_{2\sigma}$ for $z<z_{rei}$.}
\label{fig:1}
\end{figure*} 

\subsection{Model parameters}  
The model has six free parameters: $\epsilon_*$, $\epsilon_w$,
  $Z_{cr}$, $m_{PopIII}$, $t_{inf}$ and $\alpha$. 
We use the observed global properties of the MW in order to
fix $\epsilon_w$ and $\epsilon_*$. To do so we compare
the results of the simulations at redshift $z=0$ with: (i) the gas
metallicity $Z_{gas}\sim \Zsun$; (ii) the stellar metallicity
$Z_{*}\sim \Zsun$; (iii) the stellar mass $M_{*}\sim 6\times
10^{10}\Msun$; (iv) the gas-to-stellar mass ratio $M_g/M_*=0.1$; (v)
the baryon-to-dark matter ratio $f_b=0.07$; (vi) the GM metallicity
$Z_{GM}\sim  0.25\Zsun$. The last quantity has been estimated using
the observed value for [O/H] in high-velocity clouds (Ganguly et
al. 2005) which are supposed to be gas leftover from the Galactic
collapse and currently accreting onto the disk.  
The MDF of Galactic halo stars observed by Beers \& Christlieb (2006)
is instead used in order to fix the values of $Z_{cr}$ and
$m_{PopIII}$ (Fig.~1, left panel). As we will discuss in detail in
Sec.~6.1, the additional parameters $t_{inf}$ and $\alpha$ are fixed
to match the Sculptor MDF without altering the MW properties. Our
fiducial model is characterized by the following parameters
values\footnote{The difference of $\epsilon_*$, $\epsilon_w$ values
  with respect to those found in SSF07 is a result of model
  improvements. Note however that the integrals of the star formation
  rate and the gas ejection rate from progenitor haloes remain
  unaltered.}: $\epsilon_*=1$, $\epsilon_w=0.002$,
$Z_{cr}=10^{-3.8}\Zsun$, $m_{PopIII}=200\Msun$,
$t_{inf}=t_{ff}(z_{vir})/4$ and $\alpha=10$. 
  
The comparison between the best-fitting model and the observed MDF
is shown in the left panel of Fig.~1. The model provides a good fit 
to the data; in particular, the selected $Z_{cr}$ value allows to 
reproduce the peculiar MDF cut-off although it cannot account for 
the two isolated hyper iron-poor stars ([Fe/H]$=-5.3$ and [Fe/H]$=-5.7$) 
and for the recent observations of a star with [Fe/H]$=-4.8$,
the first detection in what was previously called ``metallicity
desert'' (Christlieb 2007). These stars can only be reproduced 
assuming $Z_{cr}\leq 10^{-6}\Zsun$, at the expenses of overpredicting
the number of stars in the range $-5.3<$ [Fe/H]$<-4$, loosing the 
agreement with the observed MDF cut-off (see SSF07 for a detailed 
analysis of the dependence of the predicted MDF on $Z_{cr}$ and $m_{PopIII}$). 
We find that most of the iron-poor stars ([Fe/H]$<-2.5$) form in
haloes which originally contain gas of primordial composition but 
which accrete material from the GM, Fe-enhanced by previous SN 
explosions. The initial [Fe/H] abundance within a virializing halo 
is then fixed by the corresponding GM iron abundance at the 
virialization redshift. The evolution of GM iron and
oxygen abundance predicted by the fiducial model is also shown in
Fig.1 (middle panel).

\section{The birth environment}

Once the fiducial model has been fixed, its parameters are used to
solve the system of equations (\ref{eq:SFR})-(\ref{eq:Mz}) for {\it all}
the progenitor haloes of the MW.   
The next point to address is the selection criteria to identify dSph galaxies
among various MW progenitors. We use two criteria: the first is
based on dynamical arguments and the second on reionization. 

We want to select virializing haloes which could become dSph
satellites. Using N-body cosmological simulations, Diemand, Madau \&
Moore (2005) show that in present-day galaxies, haloes corresponding
to rare high-$\sigma(M,z)$ density peaks\footnote{The quantity
  $\sigma(M,z)$ represents the linear  
theory rms density fluctuations smoothed with a top-hat filter of mass 
$M$ at redshift $z$.} are more centrally concentrated. The probability of a
protogalactic halo to become a satellite increases if it is
associated with lower-$\sigma$ density fluctuations.  
This result, combined with the fact that at each redshift $95\%$ of
the total dark matter lies in haloes with mass $M<M_{2\sigma}$, 
which correspond to $<2\sigma$ fluctuations, suggest that most
satellites originate from such density peaks. Therefore, we select dSph 
candidates from haloes with masses $M_{4}<M<M_{2\sigma}$. In Fig.~1 
(right panel) we show the redshift evolution of $M_4(z)$, defined in eq. 
\ref{eq:M4} as $M_4(z)=10^{8}\Msun [(1+z)/10]^{-3/2}$, and of the halo masses 
corresponding to $1-3$ $\sigma(M,z)$ density peaks. Note that, in addition, 
the adopted dynamical criterion can be used to set an upper limit to the dSph 
candidates formation epoch of $z_{vir}<9$ (see Fig.~1, right panel). 

The second criterion is based on reionization. During this epoch, 
the increase of the Inter Galactic Medium (IGM) temperature causes the
growth of the Jeans mass and consequent suppression of gas infall in low-mass
objects. In particular, cosmological simulations by Gnedin (2000) 
show that below a characteristic halo mass-scale the gas fraction
is drastically reduced compared to the cosmic value. We adopt
a simple prescription and assume that after reionization the
formation of galaxies with circular velocity $v_c<30$~km/s is suppressed, 
i.e. we assume that when $z<z_{rei}$ haloes with masses 
below $M_{30}(z)=M(v_c=30$~km/s~$,z)$ have no gas (for a thorough discussion 
of radiative feedback see Ciardi \& Ferrara 2005 and Schneider et al. 2007). 
Since $M_{30}(z)>M_4(z)$ (see the right panel of Fig.~1) and the
probability to form a newly virialized halo with $M>M_{30}$ is very
low, the second criterion implies that the formation of dSph candidates
is unlikely to occur below $z_{rei}$. As we will discuss in Sec.~6.4
this fact may have important consequences as, for example, the
possible existence of a universal mass of dSph host haloes.

Following the two above criteria, dSph candidates can only
form in the redshift range  $z_{rei}<z_{vir}<9$. From the middle panel
of Fig.~1 it is evident that in this redshift range the mean GM iron
abundance is $-2.5\lsim$[Fe/H]$\lsim -3$. This implies that the birth
environment of dSph candidates is pre-enriched to [Fe/H] values
consistent with those implied by the MDF observations of Helmi et
al. (2006). 

In what follows, we will present the results obtained by our fiducial
model averaged over 100 different realizations of the hierarchical merger 
tree of the MW. In each single realization, dSph candidates are selected 
from haloes with masses and redshifts corresponding to
the shaded area in the right panel of Fig.~1, that is $M_{4}<M<M_{2\sigma}$ 
for $z>z_{rei}$ and $M_{30}<M<M_{2\sigma}$ otherwise. 
Their subsequent evolution is followed in isolation with respect to the 
forming Galaxy: they neither merge nor accrete material from the GM. 
Following Choudhury \& Ferrara (2006), we vary the reionization redshift 
within the range $5.5<z_{rei}<10$. The total number of dSph candidates
depends on $z_{rei}$ and it is typically larger than the number of
observed ones. Therefore, for each $z_{rei}$, it is necessary to
randomly extract a sub-sample in order to match the total number of known MW
satellites, $\sim 15$. 
The average properties of a dSph galaxy presented in the following Sections 
refer to the case $z_{rei}=6$ (see the discussion in Sec.~6.1) and are
obtained averaging over the selected satellites from all the 100
realizations of the MW merger tree (about $\sim 2000$ objects).            

\section{Feedback-regulated evolution}

The life of a dSph is very violent in the first hundred Myr,
due to mechanical feedback effects which are more intense in 
low mass objects. The evolution of the mass of cold gas (eq.~\ref{eq:Mg}) 
helps in understanding this rapid evolution. Fig.~2 shows the evolution 
of several properties of an average dSph galaxy ($M=1.6\times 10^8 \Msun$) 
that virialized at $z_{vir}=7.2$, with respect to the formation time (age) 
$T=t-t_{vir}$. Three main evolutionary phases are identified in 
the Figure, depending on the dominant physical processes.

An increasing fraction of cold gas is collected during Phase~I
($T<40$~Myr) dominated by the infall rate. The mass of the infalling
gas rapidly increases during this epoch, reaching a maximum when 
$T=t_{inf}\sim 25$~Myr. The mass of ejected and returned gas start 
to contribute to eq. (11) only when the most massive SNe of $40\Msun$
explode\footnote{Stars with $40\Msun<m<100\Msun$ are predicted to
  collapse to black holes (Woosley \& Weaver, 1995) while massive
  Pop~III stars cannot be produced in dSph galaxies since their
  birth environment is pre-enriched to $Z_{vir}\sim 10^{-3}\Zsun >
  Z_{cr}=10^{-3.8}\Zsun$}, $\sim 6$~Myr from the formation 
of the first stellar generation. Thereafter, the mass of ejected 
and returned gas rapidly grow, due the raising number of SNe and evolving 
low mass stars. The ISM metallicity and iron abundance evolve 
accordingly during this phase: they are steadily equal to the values 
of the infalling gas ($Z_{vir}\sim 10^{-3}\Zsun$, [Fe/H]$\sim -2.9$) 
before the first SNe explodes and then rapidly increase.

\begin{figure}
  \centerline{\psfig{figure=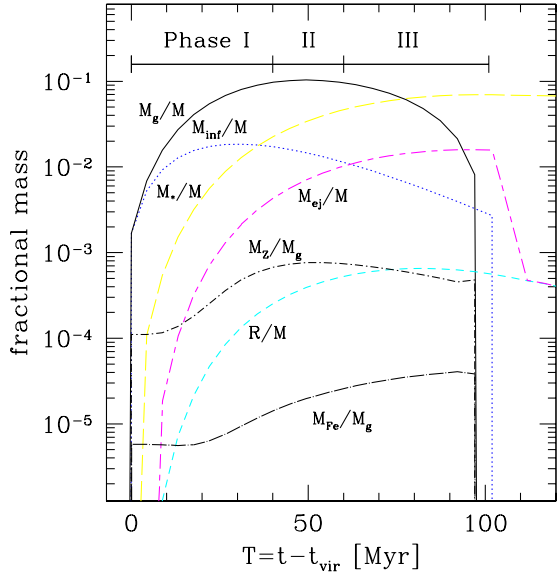,width=8.0cm,angle=0}}
\caption{Evolution of $M_g/M$ (solid line), $M_*/M$ (long dashed
  line), $M_{inf}/M$ (dotted line), $M_{ej}/M$ (long-short dashed
  line), $R/M$ (short dashed line), $M_Z/M_g$ (dotted-short dashed line),
  $M_{Fe}/M_g$ (dotted-long dashed line) for a typical dSph with total mass
  $M=1.6\times 10^8\Msun$, that virialize at redshift 
  $z_{vir}=7.2$, with respect to its age $T=t-t_{vir}$. The three main
  evolutionary phases (Phase~I $T<40$~Myr, Phase~II $40$~Myr~$<T<60$~Myr,
  Phase~III $T>60$~Myr) are also shown.}    
\label{fig:2}
\end{figure}

During Phase~II ($40$~Myr$\lsim T\lsim 60$~Myr) the gain of cold
gas by infall is mostly used to form stars and $M_g$ remains constant. 
Note that the $M_*/M$ curve in Fig. 2 represents the {\it total} stellar 
mass at time $T$. 
  
Finally, during Phase~III ($T\gsim 70$~Myr), the mass of
the ejected gas overcomes the infalling gas and $M_g$ starts to
decrease. Because of the metal-enhanced wind prescriptions $M_Z$ 
and $M_{Fe}$ should in principle decrease earlier and faster than $M_g$. 
This is the case for $M_Z$: in Fig.~2 the metallicity is a
slowly decreasing function both during Phase~II and Phase~III so that
$|\dot{M_Z}| < |\dot{M}_g|$. Conversely, the $M_{Fe}/M_g$ ratio is
enhanced during these epochs: the mass of newly synthesized iron released
by a SN with a $m=12\Msun$ progenitor is $\sim 2$ orders of magnitude bigger 
than for $m=40\Msun$ (Woosley \& Weaver, 1995)\footnote{This result is 
  virtually independent of the initial metallicity of the star. In the 
  same mass range, the total mass of metals produced remains constant.}; 
for this reason, when lower mass SNe evolve, a larger amount of iron is 
injected into the ISM and the second right-term in eq.~\ref{eq:Mz} 
can contrast the high ejection rate.       

When $T\sim 100$~Myr the mass of gas lost due to winds becomes larger
than the remaining gas mass and $M_g$ drops to zero. During this
blow-away metals and iron are also ejected out of the galaxy. Moreover, 
since SN explosions continue at subsequent times, even the infalling 
gas can rapidly acquire enough energy to escape the galaxy. The infall
is first reversed and in few Myr, when the remaining mass  
of hot gas has blowed-away, definitively stopped. 
The occurrence of reversal infall in high-redshift dwarf galaxies 
is confirmed by numerical simulations (Fujita et al. 2004). Eventually at 
$T \sim 100$Myr our template dSph is a gas free system. 
\section{Beyond blow-away}

\begin{figure}
  \centerline{\psfig{figure=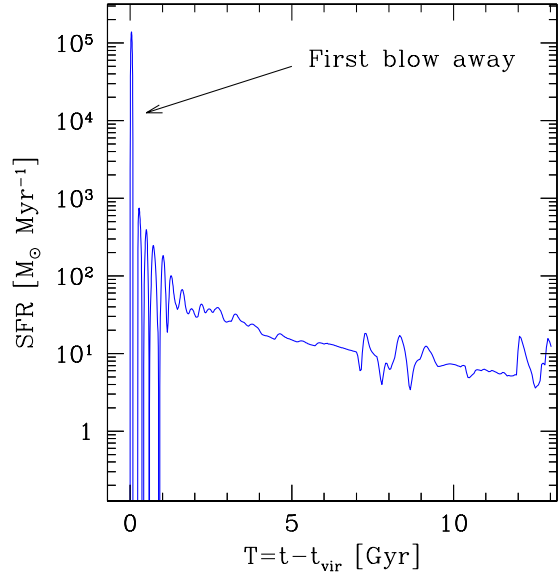,width=8.0cm,angle=0}}
\caption{SFR of a typical dSph with total mass $M=1.6\times 10^8\Msun$,
  which virialize at redshift $z_{vir}=7.2$, as a function of its age
  $T=t-t_{vir}$. The arrow shows the occurrence of the first blow-away.}
\label{fig:3}
\end{figure}

In Fig.~3 we show the star formation rate (SFR) of a typical dSph 
galaxy as a function of its age. The highest peak corresponds to the 
star formation activity during the first $100$~Myr i.e. before the blow
away. After the blow-away the galaxy remains gas free and star formation 
is suddenly halted i.e. ${\mbox SFR}(T)=0$. The gas returned by evolved stars
represents the only source of fresh gas for the galaxy after the blow
away. However, until the latest SN explodes, this low mass of gas is
easily ejected outside the galaxy by SN winds (Fig.~2); the dSph
remains dormant (${\mbox SFR}=0$) for the subsequent $\sim 150$~Myr (this
time-lag corresponds to the life time of the lowest $m=8\Msun$ SN
progenitor formed just before the blow-away). Observationally, mass
loss from evolved stars has been invoked by Carignan et al. (1998) in
order to explain the detection of neutral hydrogen (H${\rm I}$)
associated with the Sculptor dSph galaxy.   
   
After the latest SN explosion, the return rate $dR/dt$ by evolved stars
with $m<8\Msun$ becomes the only non-zero term of eq.~\ref{eq:Mg}
and the dSph enters a rejuvenation phase: the recycled gas is
collected into the galaxy and star formation starts again.

From the beginning of the rejuvenation phase the subsequent evolution
of the galaxy proceeds like in the first 100 Myr of its life. 
However (Fig.~3) SFR is now more than 2 orders of magnitude lower than
before the first blow-away, due to the paucity of returned gas. Almost
$100$~Myr later the galaxy is drained of the whole mass of gas and
metals: a new blow-away has occurred and the cycle starts again.    

In Fig.~3 we note that the repetition of blow-away and
rejuvenation phases causes an intermittent SF activity with a typical
blow-away separation of $\sim 150$ Myr. Such burst-like SFH is
similar to that inferred from the CMD observed in dSph galaxies such
as Carina (Smecker-Hane et al.~1994) although the typical duration of 
active and quiescent phases is $\sim (1-2)$ Gyr. In agreement with the 
present work, recent simulations for the collapse of an isolated 
dwarf galaxy (Stinson et al. 2007), show that feedback effects cause 
a periodic SF activity, with a typical duration of active and quiescent 
phases of $\sim 300$~Myr. 

Fig.~3 also shows that about 1~Gyr after the formation of the
galaxy, this burst-like SFH ends. Because of the gradually smaller mass
of gas returned by stars, fewer SNe are produced during this epoch. For
this reason, gas ejection is less efficient and can be easily
counteracted by the continuous input of returned gas. 

Although the SF activity continues until the present-day, the mass of
stars formed after the first blow-away ($M_*\sim 2 \times 10^5\Msun$) is only
$1\%$ of the mass of stars formed before the blow-away ($M_*\sim 2
\times 10^7 \Msun$). Consistent with this result, the analysis of the dSph 
CMD diagrams by Dolphin et al. (2005) shows that dSphs typically form 
most of their stars over 10 Gyr ago.

After the first blow-away subsequent stellar generations 
formed out of gas recycled by low mass star. The characteristic
iron-abundance of this gas is [Fe/H]$\sim -1.5$, as can be
inferred using the results of van der Hoek \& Groewengen (1997).     

\section{Observable properties}
In the following Sections we will compare our numerical results for
dSph galaxies with the most relevant observations. Given the
amount of available data we take Sculptor as the best case of a dSph
template; however, the validity of model results is general.   

\subsection{Metallicity distribution function}
In this Section we analyze the Metallicity Distribution Function (MDF)
of Sculptor, i.e. the number of relic stars as a function of their iron
abundance [Fe/H], a quantity commonly used as a metallicity tracer. 
In Fig.~4 we compare the MDF observed by Helmi et~al. (2006) with the
simulated one, normalized to the total number of observed stars (513). 
The theoretical MDF is obtained as follows: we adopt a reionization
redshift of $z_{rei}=6$; given this choice, the average number of dSph 
candidates in each realization is $N_{tot}\sim 200$, among which 
$10\%$ become MW satellites, hence naturally matching the number of
observed satellites. A higher reionization redshift of $z_{rei}=8.5$
would reduce the number of dSph candidates to $N_{tot}\sim 5$, well
below the observed value. This allows to put a solid constraint on the
reionization redshift of $z_{rei}<8.5$. As can be inferred from the
Figure, the model shows a good agreement with the observed MDF,
particularly for [Fe/H]$<-1.5$. A marginally significant deviation is
present at larger [Fe/H] values.      

We have already discussed in the previous Section that the bulk of stars
($\sim 99\%$) in a dSph galaxy is formed during the first $100$~Myr of
its life when [Fe/H]$<-1.5$. Essentially, stars formed after the first blow 
away ([Fe/H]$>-1.5$), are unnoticeable in the normalized MDF. For this
reason the physical processes regulating the MDF shape are mostly
those responsible for the cold gas mass evolution analyzed in
Sec.~4. We can use the evolution of $M_{Fe}/M_g$ shown in Fig.~2 in
order to convert time in [Fe/H] variable and identify the three main
evolutionary phases into the MDF. 

We find that stars with [Fe/H]$\lsim-2$ formed during the
infall-dominated Phase~I; the MDF shape at low [Fe/H] values is then
essentially regulated by the functional form of the infall rate. Stars
with $-2\lsim$[Fe/H]$\lsim-1.6$ (around the MDF maximum) are
formed during Phase~II i.e. when the mass of cold gas remains
approximately constant; the maximum of the MDF is instead fixed by
the values of $t_{inf}$ and $\alpha$. In particular, $t_{inf}$
determines the beginning of Phase~II and $\alpha$ its end.  
Their values ($t_{inf}=t_{ff}(z_{vir})/4$,
$\alpha=10$) have been selected in order to match the Sculptor MDF
maximum/shape without altering the global MW properties and Galactic
halo MDF (the same parameter are in fact applied to all the virialized
MW building blocks). Finally, stars with [Fe/H]$\gsim -1.6$, are
formed during the feedback-dominated Phase~III. Note in particular
that the value of the MDF cut-off ([Fe/H]$\sim -1.5$) corresponds to
the gas iron-abundance at the blow-away. 

At [Fe/H]$\gsim -1.5$ our model slightly underpredicts the data
as the theoretical MDF drops very steeply. The explanation for 
such disagreement is likely to reside in our simplified  dynamical treatment 
of mechanical feedback. Interestingly, Mori, Ferrara \& Madau (1999),
investigated the dynamics of SN-driven bubbles in haloes with $M=10^8
\Msun$ at $z=9$ using 3D simulations. They found that less than $30\%$
of the available SN energy gets converted into kinetic energy of the
blown away material, the remainder being radiated away. A large fraction of
gas remains bound to the galaxy, but is not available to form stars
before it cools and rains back onto the galaxy after $\sim 200$~Myr.
Such effect is not included in our modeling. Qualitatively we do 
expect that such ``galactic fountain" would increase the amount of
Fe-enriched gas to restart SF after blow-away, and hence the number of
[Fe/H]$\geq -1.5$ stars.

The total number of relics stars shown in the MDF corresponds to a
total stellar mass of $M_*= (3\pm 0.7) \times 10^6 \Msun$.
Using the total (dark+baryonic) dSph mass, derived from our simulations
$M=(1.6\pm 0.3)\times 10^8 \Msun$ we can compute the
mass-to-luminosity ratio     
\be
\left(\frac{M}{L_*}\right)=\left(\frac{M}{M_{*}}\right)\times
\left(\frac{M_*}{L_*}\right)\sim 150 \;\;\; ,
 \label{eq:M_L}
\ee 
having assumed $(M_*/L_*)=3$, in agreement with the results by
Ricotti \& Gnedin 2005. This result is consistent with the 
most recent estimate for Sculptor (Battaglia 2007; Battaglia
et~al. 2008), that gives a very high value $(M/L)=158\pm 33$. 

\begin{figure}
  \centerline{\psfig{figure=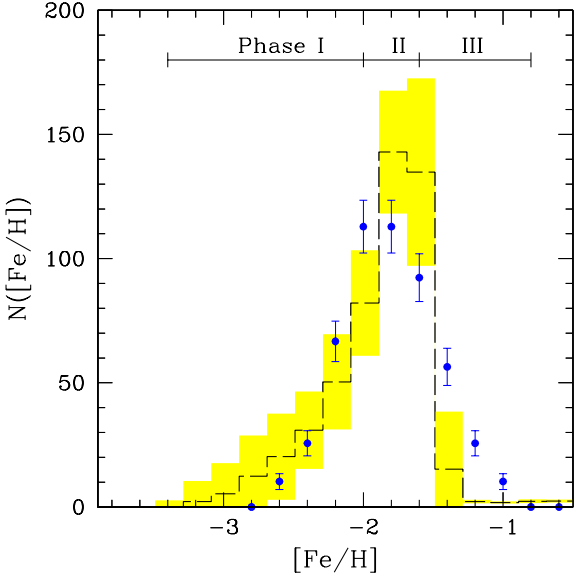,width=8.0cm,angle=0}}
\caption{Comparison between the Sculptor MDF observed by Helmi et
  al. (2006), (points) and simulated one obtained by assuming
  $t_{inf}=t_{ff}/4$, $\alpha =10$, $z_{rei}=6$ (histogram). Error
  bars are the Poissonian errors. The histogram is the averaged
  dSph MDF over the surviving satellites ($\sim 20$) in
  all the 100 realization of the merger tree ($\sim 2000$ objects). The
  shaded area represents the $\pm 1\sigma$ Poissonian error.}
\label{fig:4}
\end{figure}
\subsection{Color-magnitude diagram}
Another comparison with data can be done in terms of color-magnitude
diagram (CMD) of the Sculptor stellar population observed by 
Tolstoy et al. (2004). CMD represents one of the best tools to study the star
formation history of a galaxy. Starting from our numerical results for
a typical dSph, we have computed the corresponding synthetic CMD using the
publicly available IAC-STAR code by Aparicio \& Gallart
(2004). Given the IMF, the SFR and the ISM metallicity
evolution, IAC-STAR allows to calculate several properties of the
relic stellar population and, in particular, the stellar magnitudes. 
We have used the stellar evolution library by Bertelli (1994) and the
bolometric correction library by Lejeune et al. (1997). Note that the
IAC-STAR input parameters for the ISM metallicity evolution must be
$Z^{IAC-STAR}>0.005 \Zsun$. No binary stars have been included. 

We adopt a randomization procedure in order to simulate the
observational errors in the synthetic CMD and compare numerical
results with data. To this aim, we first derive the normalized
error distribution for the magnitude $M_I$ and the color index $V-I$ 
from the data sample by Tolstoy (private communication).   
Errors have been randomly assigned at every synthetic star, identified
by a $(M_I$, $V-I)$ pair using a Monte Carlo method and randomly
added or subtracted. Note that more accurate (and complicated)
randomization procedures exist (see for example Aparicio \& Gallart
2004); however, we consider the simple approach adopted here adequate
for our present purposes. In Fig.~5 we compare the synthetic and
observed CMDs. Data by Tolstoy et al. ($\sim 10300$ stars into the
relevant $M_I$, $V-I$ range) have been normalized to the total number
of synthetic stars derived by IAC-STAR ($\sim 2300$). In order to do
so stars have been randomly selected from the data sample.

\begin{figure}
  \centerline{\psfig{figure=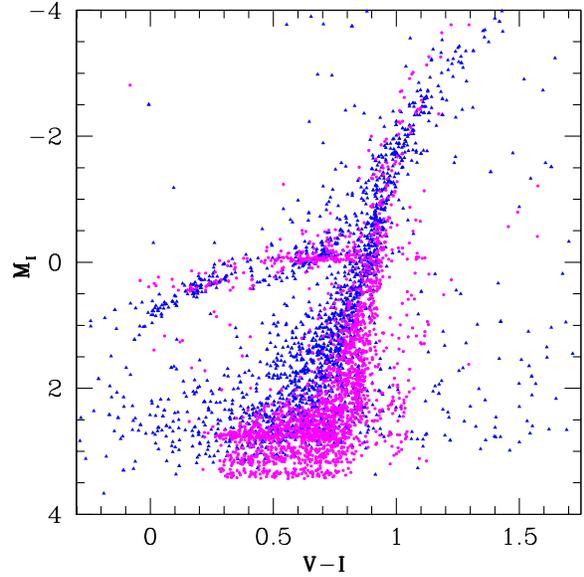,width=8.0cm,angle=0}}
\caption{Comparison between the CMD of the Sculptor stellar population
observed by Tolstoy et al. (2004) (triangles) and the synthetic CMD
(open points) derived for a typical dSph galaxy with total mass $M=1.6\times
10^8\Msun$ which virialize at redshift $z_{vir}=7.2$.}
\label{fig:5}
\end{figure}

The match between theoretical and experimental points is quite good. 
We note however that the number of red giant branch (RGB) stars in 
the synthetic CMD is lower than the observed one. 
This discrepancy can be explained with the contamination of the data 
sample by Galactic foreground stars (see Tolstoy et
al.~2004). The synthetic CMD reproduces reasonably well 
the blue/red horizontal branch stars (BHB/RHB stars)
i.e. stars residing in the CMD branch ($0< M_I< 1$, $0<(V-I)<1$). A
well populated HB in the CMD diagram might be interpreted as an indication 
of an old stellar population (age $>10$ Gyr).
The interpretation of the blue and red HB, on the contrary, is quite
controversial: due to the age-metallicity degeneracy of the CMD
stellar colors become bluer when stars are younger and/or poorer in
metallicity. For this reason the position of a star in the CMD cannot
be unequivocally interpreted.
In our model the majority of the stars are formed during the first
100~Myr of the dSph life; this means that all the stars have basically
the same age $\gsim 13$~Gyr; so the HB morphology, in our model,
reflects the metallicity gradient of the stellar populations: BHB
stars belong to metal-poor stars formed during the Phase~I (see
Fig.~2) while RHB stars to the more metal-rich stars formed during the
Phase~II.    

\subsection{Key abundance ratios}

A method commonly used to break the age-metallicity degeneracy and
derive accurate SFH from the CMD diagrams, is the analysis of the
stellar elemental abundances. In most of the observed dSph
galaxies the abundance ratio of $\alpha$ elements (O, Mg, Si, Ca)
relative to iron ([$\alpha$/Fe]) shows a strong decrease when
[Fe/H]$> -2$ (Venn et al. 2004). Since $\alpha$-elements are
primarily generated by SN II while a substantial fraction of iron-peak
elements (Fe, Ni, Co) are produced by type Ia SNe (SNe~Ia), the
decline of [$\alpha$/Fe] is usually interpreted as a contribution by
SNe~Ia. Using this argument and the assumption that
the lifetime of SNe Ia is around 1.5~Gyr, Ikuta \& Arimoto (2002)
inferred an age spead of 1-2~Gyr in the dominant stellar population of
Draco, Sextans and Ursa Minor dSphs. However, the issue of the lifetime of
SNe~Ia remains quite debated and uncertain, with timescales as short
as 40 Myr having been suggested (see Ricotti \& Gnedin 2005 for a 
thorough discussion) under starburst formation conditions.

\begin{figure}
  \centerline{\psfig{figure=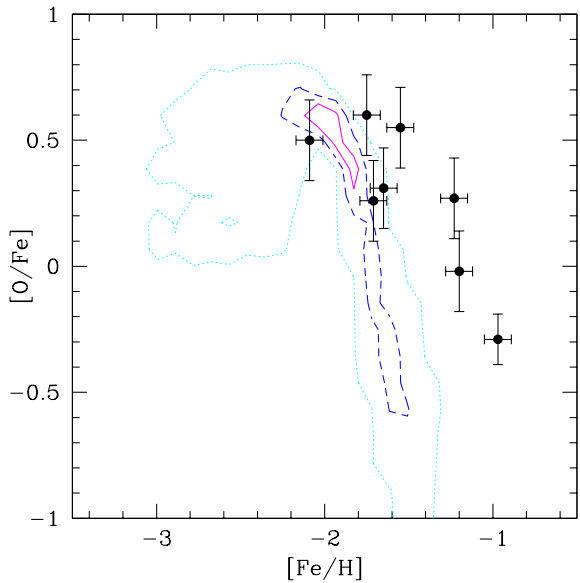,width=8.0cm,angle=0}}
\caption{Oxygen-to-iron stellar abundance with respect to [Fe/H] for
  the Sculptor dSph. Points refer to 8 Sculptor stars observed by
  Gaisler et al. (2005) and Shetrone et al. (2003); contours represent
  the probability, respectively equal to 0.6 and 0.8 and 0.99, to find
  a star into the [O/H]-[Fe/H] plane.}
\label{fig:5}
\end{figure}

In Fig.~6 we compare the oxygen-to-iron stellar abundance with respect
to [Fe/H] for 8 stars observed in Sculptor by Gaisler et al. (2005) 
and Shetrone et al. (2003), with the results of our model. In spite of
the poor statistics the data show a clear indication of the [O/Fe]
decrement for [Fe/H]$>-1.8$; in particular, subsolar values are
observed. Even if SNe Ia are not included in our model, a drop
in the [O/Fe] occurs as a result of having released the IRA
approximation (see also Fenner et al. 2006) and of differential winds. 
However, subsolar [O/Fe] values can only be accounted for by
differential winds. This is because when [O/Fe] reaches the maximum
value\footnote{The relative production rate of oxygen with respect to
  iron is larger in low-mass SNe II.}, the ``effective oxygen yield''
($dY_{O}/dt - Z^{w}_{O} dM_{ej}/dt$) is strongly reduced with
respect to iron due the effect of differential winds, which, as can be
deduced from eq.~(12), have a larger impact on more abundant elements
($Z^w_i=\alpha Z^{ISM}_i$). This causes a pronounced and rapid
decrease of [O/Fe] to subsolar values. We note that model results tend
to underpredict the observed stellar abundances: essentially,
differential winds are too efficient. We recall, however, that the
value $\alpha=10$ have been selected in order to match the Sculptor
MDF. The problem of the lack of [Fe/H]$>-1.5$ stars noted in Sec.~6.1,
is evident also here. A more sophisticated treatment of differential
winds and/or the inclusion of the missing physical effects discussed
in Sec.~6.1, should presumably remove such discrepancies. Noticeably
the result of Fig.~6 is fully consistent with the analysis by Fenner
et al. (2006), who studied the Sculptor chemical evolution including
differential winds. In conclusion we find that the trend of
[$\alpha$/Fe] does not require a prolonged star formation phase
($>1$~Gyr) but can be satisfactorily explained even if $99\%$ of the
stars formed during the first $100$~Myr of the dSph lifetime.
 
Additional constraints on the SFH may also come from the analysis of
the abundances of s-elements associated with the slow neutron-capture
process. These are produced by low-mass stars during Asymptotic Giant
Branch (AGB) phases. From an analysis of [Ba/Y] Fenner et al (2006)
concluded that most of the stars must be formed over an interval of at
least several Gyr to allow time for metal-poor AGB stars to enrich the
ISM up to the observed values. Our model does not make specific
prediction on s-elements; it is likely however that since the bulk of
stars is predicted to formed on a time-scale of $\sim100$~Myr there
would not be enough time for the ISM to be enriched with the products
of AGB stars. Nevertheless this may not be 
the only scenario to explain s-elements abundances. For example binary
systems in which the lower mass, long-living star, accretes s-enhanced
gas directly from the companion rather than from the ISM can equally
explain the observed high s-elements abundances. Internal production
during the dredge-up phase can represent yet another possibility.
These alternative scenarios are supported by the observed
stellar [s/Fe] ratio with respect to [Fe/H]. The data (Venn et
al. 2004) show that the [s/Fe] values do not increase at higher
[Fe/H], as expected if the ISM is gradually enriched by the
contribution of lower mass stars. Moreover, a large [s/Fe] spread is
observed for any [Fe/H] which is expected if the efficiencies of
accretion, dredge-up and s-element production are functions of stellar mass.

\subsection{Dark matter content} 
DSph galaxies represent the most dark matter-dominated systems known in
the Universe. It is then very interesting to determine their dark matter
mass. Observationally, the mass content of dSph galaxies is derived by
measuring the velocity dispersion profile of their stellar
populations and comparing it with the predictions from different
kinematic models. The latter step strongly depends on the adopted stellar 
kinematics (in particular the assumed velocity anisotropy radial profile), 
on the dark matter mass distribution, and on the nature of the dark matter
itself.     

Recently, Battaglia (2007), Battaglia et al. (2008)
have derived the velocity dispersion profile of Sculptor measuring the
velocities of $\sim 470$ RGB stars. They model Sculptor as a two
component system with a metal poor and a metal rich stellar population
that show  different kinematics. They use these two components as distinct
tracers of the same potential and find that the best model is a cored
profile with $r_c=0.5$kpc and $M(<r_{last})$\footnote{$M(<r_{last})$
  is the mass enclosed within the last measured point.}$=(3.4\pm
0.7)\times 10^8\Msun$ which gives an excellent representation of the
data assuming an increasing radial anisotropy. 
Interestingly, the values of $M(<r_{last})$ obtained assuming a 
NFW model for the dark matter distribution or a constant radial 
anisotropy are $M(<r_{last})=(2.4^{+1.1}_{-0.7})\times 10^8\Msun$ and
$M(<r_{last})=(3.3\pm 0.8)\times 10^8\Msun$, respectively, consistent
with the above result within 1$\sigma$.

The average mass of dSph galaxies that we infer from our simulations,
$M=(1.6\pm 0.8)\times 10^8\Msun$. Because the last measured points in the
Battaglia (2007), Battaglia et~al. (2008) typically reach 1-2 kpc, one could 
suspect that additional dark matter could be located outside this radius, 
thus turning their determination into a lower mass limit. However, for the 
mean mass value and formation redshift that we have obtained 
$M\approx 10^8 M_\odot,z_{vir}\approx 7$, the virial radius of such a halo is 
1 kpc. Thus, the agreement (at $1-2\sigma$ level) between our prediction and 
the actual mass determinations might not be coincidental, but reflects the fact 
that in these small objects star formation has propagated up to the most 
remote galactocentric regions. This prediction could be eventually checked
by deeper observations and/or other techniques. 

The narrow dispersion around the average mass value found is an indication 
that dSphs may have a universal host halo mass. This finding agrees with 
the results by Mateo et al. (1998), Gilmore et~al. (2007) and Walker et~al. 
(2007), who suggest that dSph galaxies might have a common mass scale 
$M_{0.6} = (2-7) \times 10^7 \Msun$, where $M_{0.6}$ is the dark matter mass 
within a radius of 0.6 kpc. Assuming a NFW density profile, $z_{vir}=0$ and 
a concentration parameter $c=35$ (Battaglia 2007) we found 
$M_{0.6} = 2.3\times 10^7\Msun$.
\subsection{Gas footprints of feedback}
A final comparison with data can be done in terms of the observed gas
properties. In the previous Sections we have shown that metal-enhanced 
winds driven by SN explosions play a fundamental role in determining 
the evolutionary times scales and properties of a dSph galaxy. Based on 
observations obtained with the Chandra X-Ray Observatory, 
Martin, Kobulnicky \& Heckman (2002) provide the first direct 
evidence for metal-enhanced winds from dwarf starburst galaxies. 
They have observed the hot X-ray-emitting gas around
the nearby dwarf galaxy NGC 1569 which entered in a starburst phase 
(10-20)~Myr ago. The X-ray spectrum they find presents strong
emission lines from $\alpha$-process elements, that require the wind
metallicity to be $Z^w >0.25 \Zsun$ i.e. larger than $Z^{ISM}=0.2\Zsun$,
supporting our assumption of metal enhanced winds $Z^{w}=10 Z^{ISM}$. In
particular, their best fit models predict the ratio of
$\alpha$-elements to Fe to be 2-4 times higher than the solar value;
it is then likely that the ISM is preferentially depleted in
$\alpha$-elements consistent with the findings shown in Fig.~6. We
stress that these observations confirm the idea that mechanical
feedback processes start to play a significant role in the dSph
evolution on a very short time-scale (10-20 Myr after the beginning of
the starburst phase).

Alternatively, the efficiency of mechanical feedback processes can be
tested using observations of neutral hydrogen (HI). The
Local Group dSph galaxies are all relatively HI poor (Mateo
1998) suggesting that little gas has remained after the main SF
phase. Within the known dSph galaxies, Sculptor is one 
of the few with detectable HI emission. Using radio observation, Carignan et
al.~(1998) derived a lower limit for the HI mass of $M_{HI} >
3\times 10^4\Msun$. Our simulation predicts an average mass of gas 
$M_g=(2.68\pm 0.97)\times 10^4\Msun$, in very good agreement with the
observed value if indeed this gas is in neutral form. According to our
model, the HI mass detected in Sculptor can be associated to gas
returned by evolved stars, an explanation also offered by Carignan et
al.~(1998). 

\section{Summary and Discussion}
We have proposed a global scenario for the formation and evolution of
dSph galaxies, satellites of the MW, in their cosmological context
using an improved version of the semi-analytical code GAMETE (GAlaxy
Merger Tree \& Evolution, SSF07). This approach allows to follow
self-consistently the dSph evolution and the MW formation and match,
simultaneously, most of their observed properties. In this context
dSphs formed within the Galactic environment, whose metallicity
evolution depends on the history of star formation and mechanical
feedback along the build-up of the Galaxy. The star formation and
mechanical feedback efficiencies of dSphs are assumed to be the same
as for all the Galactic building blocks; they are calibrated to
reproduce the observable properties of the MW.

DSph candidates are selected among the MW progenitors following
a dynamical and a reionization criteria; we choose haloes with masses
(i) $M_4 <M< M_{2\sigma}$ if $z>z_{rei}$, (ii) $M_{30}<M<M_{2\sigma}$ if
$z<z_{rei}$ i.e. we assumed the formation of galaxies with circular
velocity $v_c<30$~km/s to be suppressed after reionization, where
$5.5<z_{rei}<10$ (Choudhury \& Ferrara 2006).  
As the number of dSph candidates found varies with $z_{rei}$, we 
determine the fraction that will become MW satellites requiring that their
number matches the observed one ($\sim 15$). Once formed, dSphs are
assumed to evolve in isolation with respect to the merging/accreting
Galaxy. In this work we present the results obtained assuming
$z_{rei}=6$. This value provides a good agreement between the Sculptor
MDF and the simulated one and gives a total number of dSph candidates
of $N_{tot} \sim 200$; hence, we suppose that $\sim 10\%$ of them
become MW satellites. 

The results of our model, supported by the comparison with observational data
and previous theoretical studies, allow to sketch a possible evolutionary
scenario for dSphs. In our picture dSph galaxies are associated with Galactic 
progenitors corresponding to low-sigma density fluctuations 
($M_4<M<M_{2\sigma}$), that virialize from the MW environment before
the end of reionization, typically when $z=7.2\pm 0.7$. Their total
(dark+baryonic) mass results to be $M=(1.6 \pm 0.7) \times 10^8\Msun$.  

At the virialization epoch the dSph birth environment is naturally 
pre-enriched due to previous SN explosions up to [Fe/H]$_{GM}\gsim
-3$, a value fully consistent with that inferred from observations
by Helmi et al. (2006). The subsequent dSph evolution is strongly
regulated by mechanical feedback effects, more intense in low mass
objects (MacLow \& Ferrara 1999). We take winds driven by SN
explosions to be metal enhanced ($Z^w=10 Z^{ISM}$) as also confirmed
by numerical simulations (Fujita et al. 2004) and by the X-ray
observations of the starburst galaxy NGC1569 (Martin et al. 2002). 
Typically, $\sim 100$ Myr after the virialization
epoch a complete blow-away of the gas caused by mechanical feedback is
predicted. The $99\%$ of the present-day stellar mass, $M_*= (3\pm
0.7) \times 10^6 \Msun$, is expected to form during 
the first 100~Myr. The stellar content of dSphs is then dominated by an 
ancient stellar population ($>13$ Gyr old), consistent with the
analysis of the dSph CMD diagrams by Dolphin et al. (2005).  

After the blow-away the galaxy remains gas-free and SF is
stopped. Fresh gas returned by evolved stars allows to restart the SF
$\sim 150$ Myr after the blow-away. The SFR, however, is drastically
reduced due to the paucity of the returned gas. Mass loss from evolved
stars has been also invoked by Carignan et al. (1998) to explain the
detection of HI in the Sculptor dSph. About $\sim 100$ Myr later, a
second blow-away occurs and the cycle starts again. Such intermittent SF
activity is similar to those observed in Carina by Smecker-Hane et
al. (1994) and to the one derived by Stinson et al. (2007) using numerical 
simulations. Roughly 1 Gyr after the virialization this burts-like SFH 
ends while the SF activity proceeds until the present-day with a rapidly 
decreasing rate. At $z=0$ the dSph gas content is 
$M_g=(2.68\pm 0.97)\times 10^4 \Msun$. 

Our model allows to match several observed properties of Sculptor dSph: 

\begin{itemize}

\item The Metallicity Distribution Function (Helmi et al. 2006).
  The pre-enrichment of the dSph birth environment accounts for the
  lack of observed stars with [Fe/H]$<-3$, a striking and common
  feature of the four dSph galaxies observed by Helmi et al. (2006).   
\item The stellar Color Magnitude Diagram (Tolstoy et al. 2003)
  and the decrement of the stellar [O/Fe] abundance ratio
  for [Fe/H]$>-1.5$ (Gaisler et al. 2005, Shetrone et al. 2003).
  The agreement found between models and observations support
  the SFH we have predicted.  
\item The DM content $M=(3.4\pm 0.7)\times 10^8 \Msun$ and the high
  mass-to-light ratio $(M/L)=158\pm 33$ recently derived by Battaglia
  (2007), Battaglia et~al. (2008); we find $(M/L)\sim
  150$ using the predicted dark  matter to stellar mass ratio and
  assuming $(M/L)_*=3$.
\item The HI gas mass content. The value derived by radio observations
  ($M_{HI} > 3\times 10^4\Msun$, Carignan et al. 1998) is in agreement
  with our findings.   
\end{itemize}

Interestingly, the model can also be used to put an upper limit on
epoch of reionization, $z_{rei}<8.5$. The total number of selected
dSph candidates in fact, is reduced below the observed one
($N_{tot}\sim 5$) if $z_{rei}=8.5$. In addition, the imprint of
reionization lies in the suppression of dSphs formation below the
reionization redshift, $z_{rei}=6$. This result is fully consistent
with the presence of an ancient stellar population ($>13$~Gyr old) in
{\it all} the observed dSph galaxies (Grebel \& Gallagher 2004).

Despite the success of the model in producing a coherent physical
scenario for the formation of dSphs in their cosmological context and
matching several of the Sculptor and MW properties, several aspects deserve a
closer inspection. Although Sculptor represents the best
template to compare with because of its average properties and the large
amount of available data, examples of deviations are already known.  
In particular the SFHs differ considerably among dSph galaxies (Dolphin et
al. 2005, Grebel \& Gallagher 2004). The Fornax CMD diagram, for
example, indicates a massive presence of younger stars than in other
dSphs (Dolphin et al. 2005; Stetson et~al. 1998; Buonanno
  et~al. 1999); the peculiarity of this object is also evident in
the observed MDF, which is a monotonically increasing function up to
[Fe/H]$\sim -1$ ( Pont et~al. 2004; Helmi et~al. 2006; Battaglia
  et~al. 2006). The dSph properties inferred in our model, including
the SFH and the MDF, are instead ``Universal''. This is a consequence
of the selection criteria, that gives a Universal dSph host halo mass, and 
of the assumed cosmological gas fraction in all virializing
haloes. Since $M_4(z)<M_{2\sigma}$ only for $z<9$ (see Fig.~1, right
panel) and the typical mass of newly virializing halo is $\sim
M_4(z)<M_{30}$, dSphs are forced to form in the redshift range 
$6 < z_{rei} <9$. Due to the small variation of $M_4(z)$ in such a
range, the dSph dark matter content is very similar in all objects and
equal to $\sim 10^8 \Msun$. 

A refinement of the reionization criterion should allow larger deviations
from the average evolutionary trend without altering it and
possibly allow reionization imprints to emerge in
their SFHs. Cosmological simulations by Gnedin (2000) show in fact that after
reionization the gas fraction in haloes below a characteristic
mass-scale is gradually reduced compared to the cosmic value. 
If included in our model such prescription might allow to form, just after the
reionization epoch, massive dSphs ($M\geq M_{4}(z=6)$) with a lower
initial gas-to-dark matter ratio. Since mechanical feedback depends on
the available mass of gas powering the SF and on the halo binding
energy, this should translate in a less efficient mechanical feedback
and a consequent more regular SF activity. In particular, if 
blow-aways do not occur, the SF could proceed until the present day
with a higher rate, allowing massive formation of younger and more
Fe-rich stars. Alternatively, random episodes of mass accretion and/or merging
with primordial composition haloes, can be invoked as external gas
sources to power the SF at lower redshifts. 
Another physical mechanism, tidal stripping by the gravitational
field of the Galaxy, might be invoked (Ibata et~al. (2001); Mayer et~al. 
(2004, 2006)) to explain the paucity of remnant gas in dSph, 
and perhaps as a mechanism of star formation suppression. In our model, 
we see no need to resort to such effect as the large majority of the gas is
expelled by SN feedback within the first 100 Myr of dSph evolution
(we recall that the amount of newly born stars after that time is
only $\approx 1$\% of the final stellar mass). Hence, by the time
dSphs find themself embedded in the MW gravitational potential,
there is little gas left to be stripped.

Finally, we like to comment on some of the model assumptions that
could affect the results of the present work. The most relevant one
certainly resides in the perfect mixing approximation which determines
the metallicity evolution of the dSph birth environment and therefore
of the low [Fe/H] tail of the dSph MDF. Our persisting ignorance on
all the physical effects regulating such process cannot allow to
improve this assumption at the moment. The problem, however, is
partially alleviated by the spread of the GM metallicity evolution 
(Fig.~1, middle panel) induced by the stochastic nature of the merger
histories, which appears to be similar to what found by sophisticated
numerical simulations of mixing in individual galaxies (Mori \& Umemura
2006). DSph galaxies formed out of a metal-poor birth environment
[Fe/H]$<-3$ are found in our model; however, their number is very small and
their statistical impact on the average MDF negligible.
Finally, the assumed PopIII IMF (here a $\delta$-function centered in
$m_{PopIII}=200\Msun$) might in principle affect the chemical
evolution of the MW environment given the large iron production of
massive stars (see also SSF07 for a more detailed description). 
A comparison of results obtained using PopIII masses in the range
$(140-260)\Msun$ and a Larson IMF, shows that the GM evolution is
independent of the assumed PopIII IMF below $z=10$. The dSph 
birth environment is therefore not affected by this hypothesis. We refer 
the reader to SSF07 for a detailed analysis of the PopIII IMF impact
on the Galactic halo MDF.

\section*{Acknowledgements}
We are grateful to G. Battaglia, N. Gnedin, E. Grebel, A. Helmi, E. Tolstoy and
K. Venn for providing us their data and enlightening discussions.
This work has made use of the IAC-STAR Synthetic CMD computation
code. IAC-STAR is supported and maintained by the computer division of
the Instituto de Astrofísica de Canarias. We are grateful to
DAVID\footnote{{\tt
    www.arcetri.astro.it/science/cosmology/index.html}} members for
discussions.  
\bibliographystyle{mn}
\bibliography{biblio}

\label{lastpage}

\end{document}